\documentclass[prd,nofootinbib,preprint,floatfix]{revtex4}

\usepackage{amsmath,amssymb}
\usepackage[normalem]{ulem}
\usepackage{graphicx}
\usepackage{array}

\newcolumntype{C}{>{~$}c<{$~}}
\newcolumntype{R}{>{~$}r<{$~}}

\preprint{\vbox{%
\hbox{\bf YITP-SB-06-52}
\hbox{\bf UB-ECM-PF-06-38}
\hbox{\bf MADPH-06-1474}}}

\begin{document}

\vspace*{.25in}

\title{Gamma Ray Burst Neutrinos Probing Quantum Gravity}
\author{M.C.~Gonzalez-Garcia}
\email{concha@insti.physics.sunysb.edu}
\affiliation{%
Instituci\'o Catalana de Recerca i Estudis Avan\c{c}ats (ICREA) \\
Departament d'Estructura i Constituents de la Mat\`eria\\
Universitat de Barcelona, 647 Diagonal, 08028 Barcelona, Spain \\
and\\
  C.N.~Yang Institute for Theoretical Physics,
  SUNY at Stony Brook, Stony Brook, NY 11794-3840, USA}
\author{Francis Halzen}
\email{flhalzen@facstaff.wisc.edu}
\affiliation{%
Department of Physics, University of Wisconsin, 
Madison, WI 53706,USA
\vspace*{.25in}}

\begin{abstract}
Very high energy, short wavelength, neutrinos may interact with the
space-time foam predicted by theories of quantum gravity. They would
propagate like light through a crystal lattice and be delayed, with
the delay depending on the energy. This will appear to the observer as
a violation of Lorenz invariance. Back of the envelope calculations
imply that observations of neutrinos produced by gamma ray bursts may
reach Planck-scale sensitivity. We revisit the problem considering two
essential complications: the imprecise timing of the neutrinos
associated with their poorly understood production mechanism in the
source and the indirect nature of their energy measurement made by
high energy neutrino telescopes.
\end{abstract}

\maketitle

\section{Introduction}
It has been realized for some time that particle physics experiments
using cosmic beams may be sensitive to Planck scale
physics\cite{icehep}. Neutrino telescopes in particular have unmatched
sensitivity to violations of Lorenz invariance (LIVs) and the
equivalence principle by exploiting the high statistics observations
of the guaranteed high energy atmospheric neutrino beam in
conjunctions with methods demonstrated by the Superkamiokande and
Macro collaborations\cite{GHM}. LIVs are the sources of observable
flavor oscillations caused by the different speed of light of the
three neutrino flavors. While the atmospheric neutrino data will reach
PeV energy, the observation of cosmic beams may extend the neutrino
sample to EeV energy. Of particular interest in this context is the
observation of high energy neutrinos from gamma ray bursts (GRBs) whose
short-time burst nature make more direct tests of Lorenz invariance
possible. Their energy is such that the measurements may be sensitive
to modifications of Lorenz symmetry associated with Planck scale
physics\cite{amelino,Choubey,piran}.

Lorenz invariance can be tested by measuring the travel times of
particles over cosmic distances. The method is familiar from pulsar
observation\cite{kaaret}. The pulsar provides a clock at the time of
emission. It has recently been suggested that more sensitive
observations can be made by measuring the relative time of photon and
very high energy neutrino emission by GRBs~\cite{piran}. 
The photon emission in a burst lasting seconds
starts the clock for a neutrino whose early or delayed arrival time
$\Delta t$ will be a signature of LIV. Note that at the energies
relevant to this problem neutrinos are, like photons, massless
particles.

The measurement conceptually tests, in an essentially
model-independent way, LIV associated with the possible breakdown of
Lorenz symmetry at the Planck scale. In some theories unifying quantum
mechanics and gravity the continuous spacetime of general relativity
becomes quantized in volumes that cannot be further
subdivided. Particles with very high energy and small wavelength may
sense this spacetime foam and be delayed, just like photons traveling
through a crystal. The phenomenology can be discussed as a generic
modification of the dispersion relation
\begin{equation}
E^2=p^2+m^2 \pm E^2({E \over \xi E_{Planck}})^n \pm ...
\end{equation}
for particles with energy $E << \xi E_{Planck}$. The time delay of a
particle with energy $E$ emitted at redshift $z$ (distance $d$) is 
\begin{equation}
\Delta t_{LIV} =\frac{1+n}{2}({E \over \xi E_{Planck}})^n 
\frac{1}{H_0}\int_0^z (1+z')^n \frac{dz'}
{\sqrt{\Omega_m(1+z')^3+\Omega_\Lambda}}
\simeq {1+n \over 2} ({d \over c})({E \over \xi E_{Planck}})^n\; .
\label{eq:delt}
\end{equation}
A sensitive test requires good timing $\Delta t$, high energy $E$ and large
distances $d$. Back-of-the-envelope is sufficient to conclude that GRBs
provide the opportunity to reach the Planck scale, i.e $\xi \sim 1$,
at least for n=1, although there is little sensitivity for
higher values of n. It is also misleading.

As will become clear as the arguments in this paper develop, it will
be challenging to ever argue that the delay of a high energy neutrino
relative to the light of a GRB requires LIV rather than a
straightforward astrophysics explanation associated with the poorly
understood nature of the source. We will argue indeed that the
fireball phenomenology of GRBs can accommodate the emission of high
energy neutrinos over timescales from seconds to days and
beyond. Emission over such timescales is actually expected, as we will
argue further on. In particular, we will give an example in the next
section how delays of very high energy neutrinos by more than 1,000
seconds from the start of the photon display, can be accommodated by
fireball phenomenology rather than by LIV. This renders the
implementation of the proposal in reference\cite{piran} more
challenging.

We here also revisit an alternative proposal to exploit the emission
of two neutrinos from a single GRB\cite{amelino,Choubey}. Even though
the neutrino rate from GRBs is expected to be order 10 per
kilometer squared per year\cite{hh}, events with two neutrinos are not
rare\cite{fluct}. This is because neutrinos are mostly produced by
relatively nearby and relatively energetic bursts. As an example we
refer to the calculation of reference\cite{guetta}. The large
burst-to-burst fluctuations can be seen from their figure 2. Although
their conservative estimates predict less than 10 events per year,
applying Poisson statistics we anticipate that bursts with 2 neutrinos
should be observed within 2 year intervals.

We will emphasize how GRB fireball phenomenology anticipates a very
attractive timing of the neutrino emission, with TeV neutrinos
produced before, PeV neutrinos in coincidence with, and EeV neutrinos
after the photon display. I.e. higher energy particles come later and
a reversal of this order can provide us with evidence for LIV.

We finally confront the additional problem that kilometer-scale
neutrino telescopes do not measure neutrino energy directly. For the
dominant $\nu_{\mu}$ signal the detector only measures the energy-loss
of the secondary muon when it passes through the
detector\cite{Halzen}. We will derive the probability distribution of
neutrino energies that accommodates the observed energy in the
detector. Our results on energy measurement presented
should be of interest beyond their application to the problem
considered here. The problem is less severe for electron and tau neutrinos
where the detectors are total absorption calorimeters for the
secondary showers produced by the neutrino
interactions\cite{Halzen}. 

In the end we conclude that revealing Planck-scale physics with
IceCube may be possible, but will be challenging. Reading the time of
the astrophysical GRB clock is not straightforward and the energy of
the neutrinos is not directly measured. We point out scenarios where
the first problem can be overcome, in the meantime we will just have
to wait for observations that will clearly define the dilemma, if
any. The second problem is interesting and impacts all observations
with high energy neutrino telescopes. A straightforward solution is to
do the science with electron and tau neutrinos only. The energy of
their secondary showers can be measured with a linear energy
resolution of 20\%. Though GRBs represent a background-free signal in
neutrino telescopes because of the clustering in time and in
direction, the event rates are still uncertain. They will exceed 10
events per kilometer square year, in the case where the baryon loading
in GRB fireballs accommodates the observed cosmic rays. With the low
event rates anticipated doing the Planck science with showers may not
be a practical solution and we may be forced to explore the long range
of the secondary muons in order to collect the enhanced rates expected
for muon neutrinos. Their better pointing will also facilitate
background rejection. How well one reconstructs muon energy of
individual events has to the best of our knowledge not been
systematically analyzed. Our results should be of interest beyond
their application to the problem considered here.

\section{Neutrino Timing by Gamma Ray Bursts}

We here review how GRBs produce a sequence of TeV, PeV and
EeV neutrinos separated in time. They provide a clock with later time
implying higher energy. We propose to exploit this feature for
measuring particle delays signaling LIV.

Gamma ray bursts are perhaps the best motivated sources of high-energy
neutrinos\cite{waxmanbahcall,mostlum2,hh,mostlum3}. The collapse of
massive stars to a black hole has emerged as the likely origin of the
``long" GRBs with durations of 10 seconds on average. The GRB rate is
consistent with the rate of supernovae with progenitor masses
exceeding several solar masses. In the collapse of the massive star a
fireball is produced which expands with a highly relativistic velocity
powered by radiation pressure. The fireball eventually runs into the
stellar material that is still accreting onto the black hole. If it
successfully punctures through this stellar envelope the fireball
emerges to produce the GRB display. While the energy transferred to
highly relativistic electrons is thus observed in the form of
radiation, it is a matter of speculation how much energy is
transferred to protons.

The phenomenology that successfully describes GRB observations is
that of a fireball expanding with highly relativistic velocity,
powered by radiation pressure. The observer detects boosted energies
emitted over contracted times; without this a description of the
extreme observations is impossible. The millisecond variations of the
GRB flux require an original event where a large amount of energy is
released in a very compact volume of order $100\,$km. The dynamics of
the fireball is actually reminiscent of the physics of the early
expanding universe. Initially, there is a radiation dominated soup of
leptons and photons and few baryons. It is hot enough to freely
produce electron-positron pairs. With an optical depth of order $\sim
10^{15}$, photons are trapped in the fireball. It cannot radiate. This
causes the highly relativistic expansion of the fireball powered by
radiation pressure. The fireball will expand with increasing velocity
until it becomes transparent and the radiation is released in the
display of the GRB. By this time, the expansion velocity has reached
highly relativistic values of order $\gamma300$.

With a relativistic expansion of $\gamma300$ it may be possible that
protons are accelerated to energies above $10^{20}$\,eV in the
fireball that accommodates the GRB
observations\cite{waxman2,waxman2b,waxman2c}. GRBs within a radius of
50-100 Mpc over which protons can propagate in the microwave
background, may therefore be the sources of the ultra high-energy
cosmic rays. The assumption that GRBs are the sources of the highest
energy cosmic rays does determine the energy of the fireball
baryons. Accommodating the observed cosmic ray spectrum of
extragalactic cosmic rays requires roughly equal efficiency for
conversion of fireball energy into the kinetic energy of protons and
electrons. In this scenario the production of neutrinos of
100--1000\,TeV energy in the GRB fireball is a robust prediction
because pions, and therefore neutrinos, are inevitably produced in
interactions of accelerated protons with fireball
photons\cite{waxmanbahcall,mostlum2,hh,mostlum3}.

There are several other opportunities for neutrino production
following the chain of events in the production of a GRB.

\begin{itemize}

\item TeV neutrino production in the stellar envelope $\sim$\,10 seconds
before the GRB display.

The core collapse of massive stars has emerged as the likely origin of
the ``long" GRB with durations of tens of seconds. The fireball
produced is likely to be beamed in jets along the rotation axis of the
collapsed object. The mechanism is familiar from observations of jets
associated with the central black hole in active galaxies. The jets
eventually run into the stellar material that is still accreting onto
the black hole. If the jets successfully puncture through this stellar
envelope they will emerge to produce a GRB. While the fireball
penetrates the remnant of the star, the fast particles in the tail
will catch up with the slow particles in the leading edge and collide
providing another opportunity for pion production yielding neutrinos
of tens of TeV energy. Bursts within a few hundred megaparsecs ($\sim
10$ bursts per year as well as an additional unknown number of
``invisible'' bursts from failed GRBs) may actually produce large rates
of TeV neutrino events in a kilometer-scale detector and, possibly,
observable rates in a first-generation detector such as
AMANDA\cite{beacom}. Interestingly, failed
jets that do not emerge will not produce a visible GRB but do produce
observable neutrinos\cite{waxmantev1,waxmantev2}.

\item PeV neutrino production inside the expanding fireball for a
duration of ~10 seconds coincident with the GRB.

The production of PeV neutrinos in cosmic ray producing GRB fireballs
is a robust prediction\cite{waxmanbahcall}. Neutrinos are inevitably
produced in interactions of accelerated protons with fireball
photons. As the kinetic energy in fireball protons increases with
expansion, a fraction of their energy is converted into pions once the
protons are accelerated above threshold for pion production.

\item EeV production in the interstellar medium over minutes~days
after the GRB display.

Afterglow observations show that external shocks are produced when the
GRB runs into the interstellar medium. The hadronic component in the
fireball can be rapidly depleted by means of photopion process on time
scales of $10^2 \sim 10^4$\,seconds. It has been argued that this
``aftershock" explains the accumulating evidence from recent SWIFT
observations for a second high energy component in the GRB as well as
for the sharp decline of the X-ray light curves that take place over
similar timescales\cite{dermer}. This is the phase in which the
highest energies are reached, 100\,EeV for cosmic rays and 1\,EeV for
the secondary neutrinos. The neutrino rates are predicted to be low in
IceCube but represent an opportunity for radio or acoustic detectors
such as RICE with high threshold, but also larger effective
area\cite{gorham}.

It is this last phase of neutrino production that will interfere with
the proposal of Jacob and Piran\cite{piran}. Optimistically assuming
that one will make multiple observations and that, contrary to
evidence, GRBs are somehow standard candles, one could still try to
distinguish LIV from astrophysics induced delays by establishing the
linear dependence of the time delay on the distance; see Eq.~(\ref{eq:delt}).

\end{itemize}

While the arrival time of neutrinos is measured with microsecond
accuracy, determination of the energy is a considerable
challenge. Both are required to test LIV.

\section{High Energy Neutrino Telescopes and $\nu_\mu$ Energy Measurements}

In neutrino telescopes, neutrinos are detected through the observation of \v
{C}erenkov light emitted by charged particles produced in neutrino
interactions. The neutrino induced events can be categorized as either
muon tracks or showers. Cosmic ray muons and muons from charged
current (CC) $\nu_{\mu}$ interactions are the origin of tracks.
Showers results from neutrino interactions --- $\nu_e\ \mbox{or}\,\,
\nu_\tau$ CC interactions, and neutral current (NC) interactions
initiated by all three flavors --- inside or near the detector.
Because of the large range of muons, kilometers to tens of kilometers
for the energies considered here, the effective volume of the detector
for muon neutrinos is significantly larger than the instrumented
volume.  Furthermore, the angular resolution for muon tracks is
superior to that for showers.  For example, for Icecube the angular
resolution for muon tracks $\approx 0.7^\circ$~\cite{Ahrens:2002dv} which 
allows a search window of solid angle 
$\Delta \Omega_{1^\circ \times 1^\circ} \approx 3 \times
10^{-4}$~SD, 
while for showers the angular resolution is significantly
worse.

While the large effective volume and the better background rejection
of atmospheric backgrounds because of the superior angular resolution
favors the observation of muon neutrinos, measurement of their energy
is indirect. Neutrino telescopes only measure the energy $E^{\rm dep
}_\mu$ deposited by the muon when traversing the detector which is only
indirectly related to the original neutrino energy. Particle physics
determines the probability $f(E_\nu |E^{\rm dep}_\mu)$ distribution
for the parent neutrino energy of a muon track that deposits an energy
$E^{\rm dep}_\mu$ inside the detector:
\begin{equation}
f(E_\nu |E^{\rm dep}_\mu)
=\frac{1}{\rm Norm}
\int^\infty_{l_{min}} dl\,
\int_{E^{\rm dep}_\mu}^{E_\nu} dE_\mu^{\rm fin}\,
\int_{E_\mu^{\rm fin}}^{E_\nu} dE_\mu^0\, 
\frac{d\phi_{\nu_\mu}}{dE_\nu}(E_\nu)\, 
\frac{d\sigma^\mu_{CC}}{dE_\mu^0}(E_\nu,E_\mu^0)\,
F(E^0_\mu,E_\mu^{\rm fin},l) \,A^0_{eff}.  
\label{eq:probenu}
\end{equation}
$\frac{d\phi}{dE_\nu}$ is the incident neutrino spectrum,
$\frac{d\sigma^\mu_{CC}}{dE_\mu^0}(E_\nu,E_\mu^0)$ is the differential
neutrino CC interaction cross section and $F(E^0_\mu,E_\mu^{\rm
fin},l)$ represents the probability that a muon produced with energy
$E_\mu^0$ arrives at the detector with energy $E_\mu^{\rm fin}$ after
traveling a distance $l$. $l$ is obtained by propagating the muons
through the ice taking into account energy losses from ionization,
bremsstrahlung, $e^+e^-$ pair production and nuclear
interactions~\cite{ls}.  The details of the detector are encoded in
the effective area $A^0_{eff}$ and $l_{min}$ is the minimum muon
track length required for the event to be detected.  ${\rm Norm}$ is
the normalization factor obtained from $\int_{E^{\rm dep}_\mu}^\infty
f(E_\nu |E^{\rm dep}_\mu)\, dE_ \nu \,=1$.

\begin{figure}[ht]
\includegraphics[width=\textwidth]{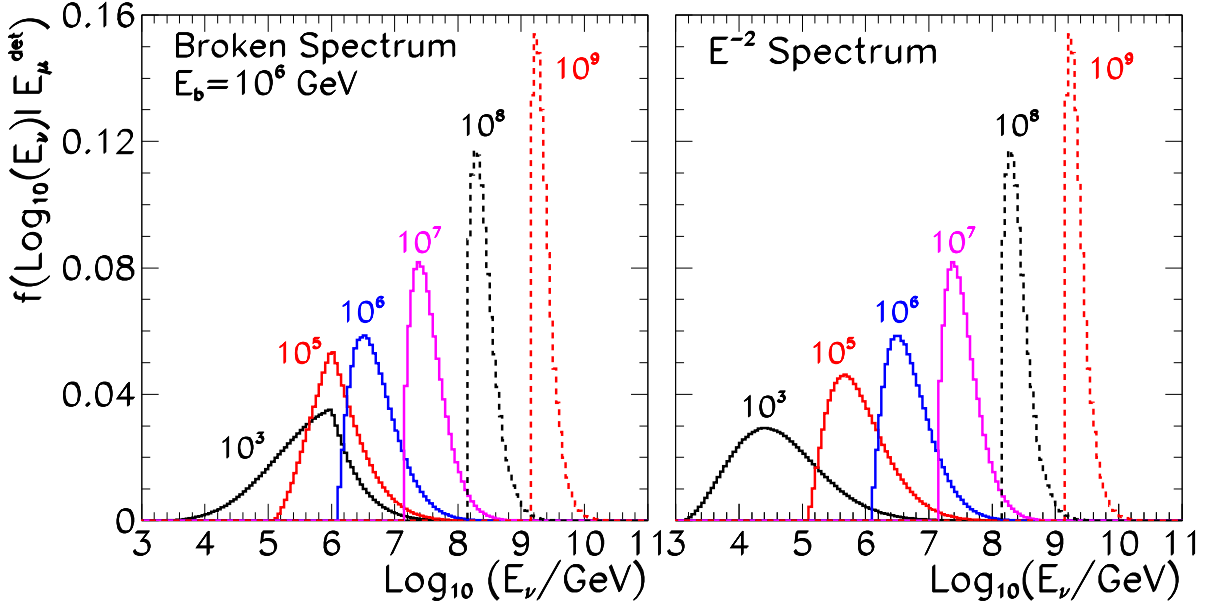}
\caption{\label{fig:penem} 
Probability distribution for the parent neutrino energy of a muon track 
event which deposits an energy  $E^{\rm dep}_\mu$ 
at the detector for different values of 
$E^{\rm dep}_\mu=10^3, 10^5, 10^6, 10^7, 10^8$, and $10^9$ GeV as labeled
in the figure. 
The left  panel was obtained  for a broken neutrino spectrum 
(Eq.(\ref{eq:broken})) with $E_b=10^6$ GeV while the right panel  
corresponds to a continuous $E_\nu^{-2}$ spectrum. }
\end{figure}

We show in Fig.\ref{fig:penem} the probability distribution function 
$f(E_\nu |E^{\rm dep}_\mu)$ ( for convenience we plot it as
$f(\log_{10}(E_\nu) |E^{\rm dep}_\mu)= 
\ln (10) E_\nu f(E_\nu |E^{\rm dep}_\mu)$) as a function of the
neutrino energy for different values of $E^{\rm dep}_\mu$ and
for two different neutrino spectra, a broken spectrum 
\begin{eqnarray}
&&\frac{d \Phi}{dE_\nu}=A\frac{1}{E_b E_\nu} \; \; \; {\rm for\; E_\nu<E_b} 
\label{eq:broken}
\\
&&\frac{d \Phi}{dE_\nu}=A\frac{1}{E_\nu^2} \; \; \;  {\rm for\; E_\nu<E_b}
\nonumber
\end{eqnarray}
and a continuous spectrum $\frac{d \Phi}{dE_\nu}=A\frac{1}{E_\nu^2}$.
We used the parametrization for $A^0_{eff}$ given in Ref.\cite{GHM}
which describes the response of the IceCube detector after all
backgrounds have been rejected (this is achieved by quality cuts
referred to as ``level 2" cuts in Ref.~\cite{IceCube}) and we fix
$l_{min}=300$ m.
 
The figure illustrates the challenge in experimentally determining the
presence of an effect which depends on the original neutrino energy
based on the observation of a muon track event.  We can furthermore
infer the even bigger challenge for the case of establishing (or
constraining) the violation of Lorentz invariance from the observation
of the arrival times of two muon events with different energy from a
single GRB, a case we discuss next.

\section{Results and Conclusions}

In principle one could test the hypothesis of 
LIV from the observed arrival times, $t_1$ and $t_2$,  of two neutrino 
events from a single GRB by confronting
\begin{equation}
\Delta t_{obs}=t_2-t_1=\Delta t_{source}+ \Delta t_{2,LIV}-\Delta t_{1,LIV}
\end{equation}
with $\Delta t_{LIV}$ given in Eq.(\ref{eq:delt}). 
Here we denote by $\Delta t_{source}$ the time difference of production
of the two neutrinos {\it at the source} and not at arrival. 
It is not directly observed. Assuming we can infer $\Delta t_{source}$, 
then for $n=1$
\begin{equation}
\xi=\pm \frac{1}{\Delta t_{obs}-\Delta t_{source}} \frac{F_1(z)}{H_0} 
\frac{E_{\nu,2}-E_{\nu,1}}{E_{Planck}}
\label{eq:xi}
\end{equation}
where we have defined
\begin{equation}
F_n(z)=\frac{(1+n)}{2} \int_0^z \frac{1}
{\sqrt{\Omega_m(1+z')^3+\Omega_\Lambda}} \, .
\end{equation}
Numerically  for $\Omega_m=0.25$ and $\Omega_\Lambda=0.75$ 
$0.28<F_1(z)<2.5$ and $0.35<F_2(z)<8.5$ for $0.5<z<5$.

However, as illustrated in Fig.\ref{fig:penem}, inferring
$E_{\nu,2}-E_{\nu,1}$ from the measured values of $E^{\rm
dep}_{\mu,1}$ and $E^{\rm dep}_{\mu,2}$ of 
two muon tracks is highly
non-trivial.  For example, for $n=1$ one can determine the relevant
probability distribution of the parent neutrino energy difference
$\Delta E_\nu$ as:
\begin{equation}
f(\Delta E_\nu|E^{\rm dep}_{\mu,1},E^{\rm dep}_{\mu,2})=
\int dE_{\nu_1} \int dE_{\nu_2} 
f(E_{\nu,1} |E^{\rm dep}_{\mu,1}) f(E_{\nu,2} |E^{\rm dep}_{\mu,2}) 
\delta (E_{\nu,2}-E_{\nu,1}-\Delta E_\nu)\, .
\label{eq:fdelta}
\end{equation} 
(Equivalently one can build the corresponding probability distribution
for any combination $E^n_{\nu,1}-E^n_{\nu,2}$.)
\begin{figure}[ht]
\includegraphics[width=\textwidth]{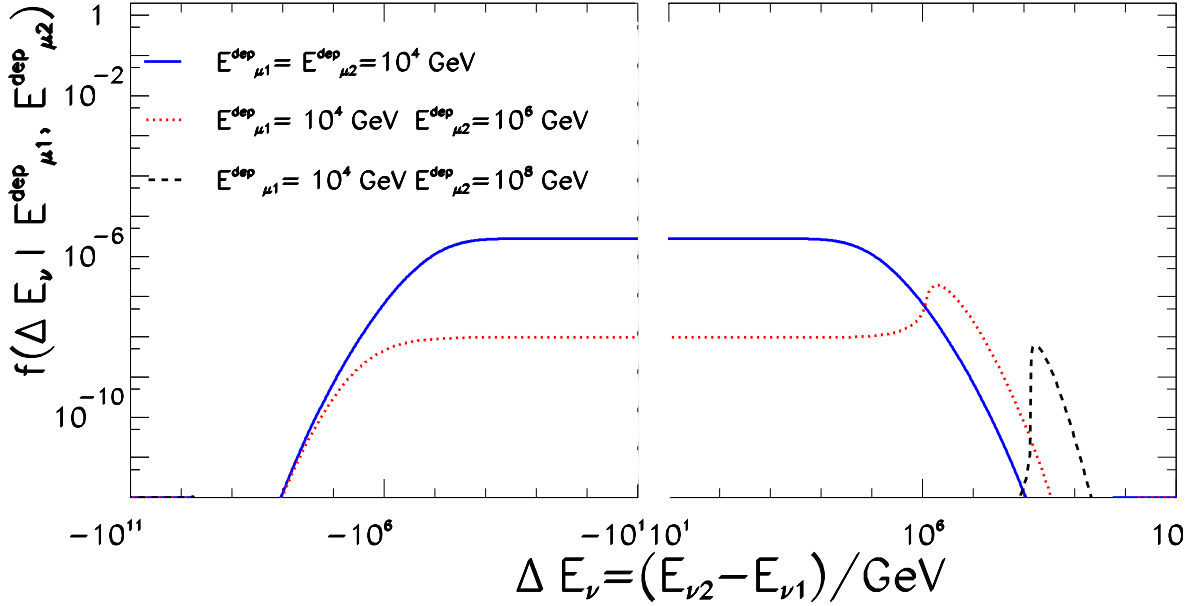}
\caption{\label{fig:fdelta} 
Probability distribution for the parent neutrino energy difference of
two muon track events which deposit an energy $E^{\rm dep}_{\mu,1}$
and $E^{\rm dep}_{\mu,2}$ inside the detector for different values of
$E^{\rm dep}_{\mu, 1}$ and $E^{\rm dep}_{\mu, 1}$ as labeled in the
figure. This distribution is for a $E^{-2}$ neutrino spectrum. In all
cases $\int_{-\infty}^{\infty} f(\Delta) d\Delta=1$ but we are showing
$f(\Delta)$ as a function of $\log(\Delta)$ hence the apparent
difference in normalization of the curves. }
\end{figure}

We show in Fig.\ref{fig:fdelta} the value of 
$f(\Delta E_\nu|E^{\rm dep}_{\mu,1},E^{\rm dep}_{\mu,2})$ for 
$E^{\rm dep}_{\mu,1}=E^{\rm dep}_{\mu,2}=10^4$ GeV, 
$E^{\rm dep}_{\mu,1}=10^4$ GeV and $E^{\rm dep}_{\mu,2}=10^6$ GeV
and $E^{\rm dep}_{\mu,1}=10^4$ GeV and $E^{\rm dep}_{\mu,2}=10^8$ GeV.
From the figure we see that, because of the relatively broad spectrum
of the parent neutrino energy distributions, one cannot rule out relatively 
large parent neutrino energy differences even if the two observed 
muon track events deposit the same energy at the detector. Conversely,
only for very different deposited energies one can establish that
the parent neutrino energies were different with high CL.

More quantitatively for a given pair of observed events with 
$E^{\rm dep}_{\mu,1}$ and $E^{\rm dep}_{\mu,2}$ 
from a single GRB we can  determine:
\begin{eqnarray}
{\rm Prob}(E_{\nu,2}-E_{\nu,1}< \Delta_{max})=
\int_{-\infty}^{\Delta_{max}} 
f(\Delta|E^{\rm dep}_{\mu,1},E^{\rm dep}_{\mu,2}) \, d\Delta   \, , \nonumber 
\\
{\rm Prob}(E_{\nu,2}-E_{\nu,1}> \Delta_{min})=
\int^{\infty}_{\Delta_{min}} 
f(\Delta|E^{\rm dep}_{\mu,1},E^{\rm dep}_{\mu,2}) \, d\Delta  \, .
\label{eq:probdel}
\end{eqnarray}
From Eqs.(\ref{eq:xi}) and (\ref{eq:probdel}) one can compute the
probability for $\xi_{max}>\xi>\xi_{min}$ which corresponds to the
observation of a pair of events from a single GRB with $E^{\rm
dep}_{\mu,1}$ and $E^{\rm dep}_{\mu,2}$ and arrival times $t_1$ and
$t_2$.

\begin{figure}[ht]
\includegraphics[width=0.7\textwidth]{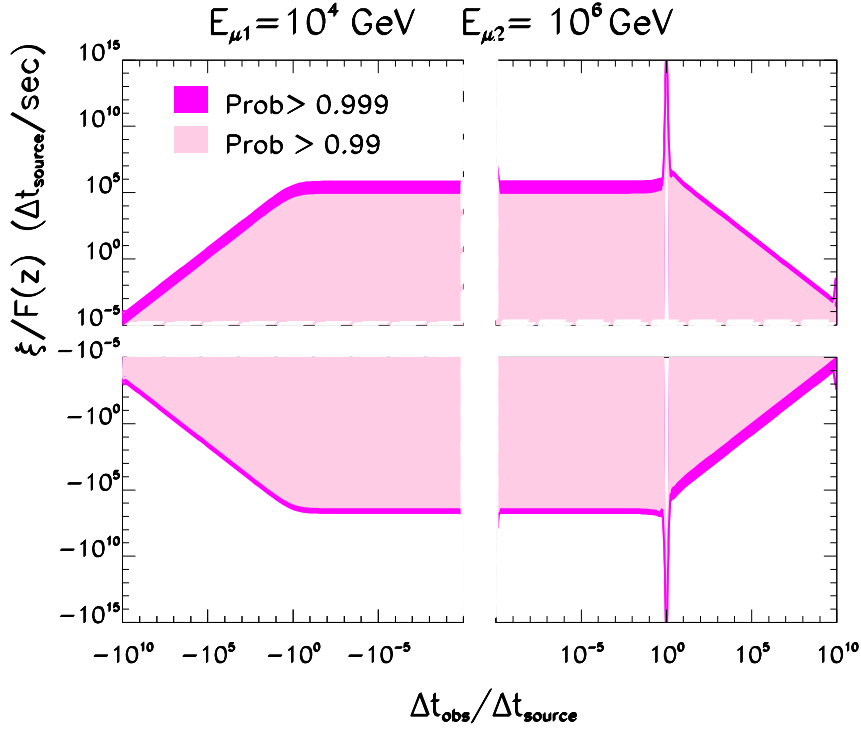}
\caption{\label{fig:con46} Allowed ranges of the LIV scale factor
$\xi$ with probability bigger than $10^{-3}$ and $10^{-2}$ from the
observation of two track events from a single GRB with $E^{\rm
dep}_{\mu,1}=10^4$ and $E^{\rm dep}_{\mu,2}=10^6$ GeV as a function of
the ratio of between the observed time difference and the time
difference at production in the GRB. This is shown for an assumed
continuous $E^{-2}$ neutrino spectrum.}
\end{figure}

\begin{figure}[ht]
\includegraphics[width=0.7\textwidth]{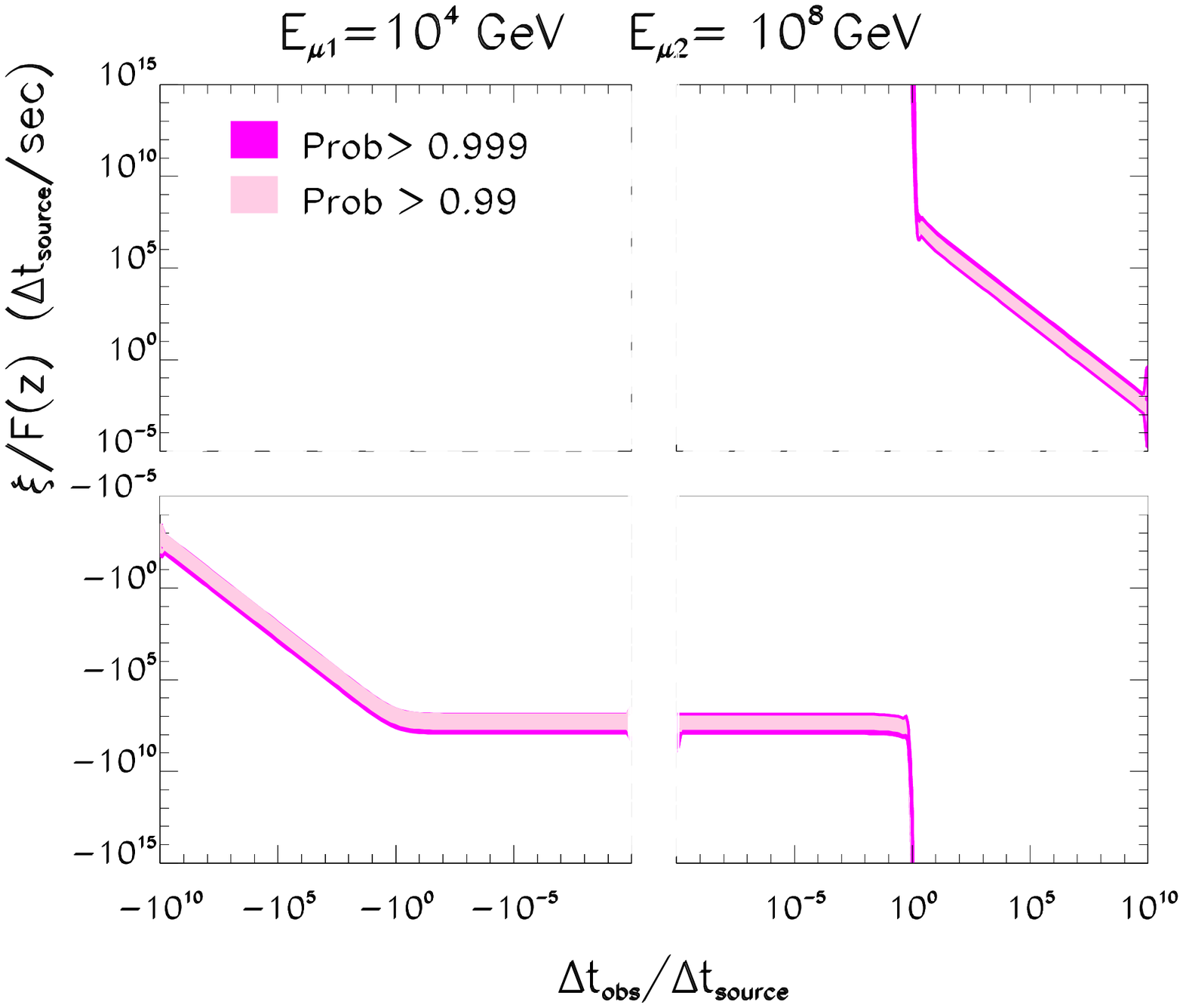}
\caption{\label{fig:con48} Same as Fig.\ref{fig:con46} but for $E^{\rm
dep}_{\mu,1}=10^4$ GeV and $E^{\rm dep}_{\mu,2}=10^8$ GeV.}
\end{figure}

As illustration we show in Figs.\ref{fig:con46} and \ref{fig:con48}
the ranges of $\xi$ which can be inferred from the observation of two
muon track events with $E^{\rm dep}_{\mu,1}=10^4$ GeV and $E^{\rm
dep}_{\mu,2}=10^6$ GeV and with $E^{\rm dep}_{\mu,1}=10^4$ GeV and
$E^{\rm dep}_{\mu,2}=10^9$ GeV respectively, as a function of the
ratio of the arrival time difference and the time difference at
production.

From the figures we see that:
\begin{itemize}
\item What one observes is in all cases compatible with no LIV if the
model predicts $\Delta t_{source}=\Delta t_{obs}$, hence the
asymptotic peaks which reach $\xi= \infty$ and $\xi=-\infty$ in the
upper and lower right panels at $\Delta t_{source}=\Delta t_{obs}$.
However, as can be seen in Fig.~\ref{fig:con46}, if the deposit energy
of the two events are not different enough, the allowed range of $\xi$
include very low values.  Thus no meaningful bound on the scale of LIV
can be set.
\item When the deposited energies of the two events are very different
and one assumes that $\Delta t_{obs} = \Delta t_{source}$, one can set
a lower bound on the scale of LIV. For example for $E^{\rm
dep}_{\mu,1}=10^4$ GeV and $E^{\rm dep}_{\mu,2}=10^8$ GeV we find at
99.9\% probability that
\begin{equation}
 |\xi|> 10^8 F_1(z) \frac{sec}{\Delta t_{source}}
\label{eq:xibound}
\end{equation}
\item In the absence of an argument for $\Delta t_{source}=\Delta
t_{obs}$ , the conclusion is that one has to be observing LIV. However,
as seen in Fig.~\ref{fig:con46}, if the two events deposit energies
which are not different enough the observation is compatible with a
very wide range of scales of LIV and with both signs of the effect.
\item 
When the deposited energies of the two events are very different the
determination of $\xi$ becomes better as implied by the narrow width
of the allowed regions in Fig.~\ref{fig:con48}.  Also one sign is
favored: it is more likely that $\xi$ has the opposite (same) sign
than $\Delta t_{source}$ if the $\Delta t_{obs} > \Delta t_{source}$
($\Delta t_{obs} < \Delta t_{source}$).
\end{itemize}

Finally, the estimates assume negligible background from atmospheric
neutrinos.  One expects of the order of 200 muon track events a year
with $E^{\rm dep}_{\mu}\geq 10^4$ GeV arriving from any direction with
zenith angle $\cos\theta<0.2$ \cite{GHM}. This means that, on average,
the expected number of atmospheric background events in an angular bin
of $\Delta \Omega_{1^\circ \times 1^\circ} \approx 3 \times
10^{-4}$~sr and in a time interval of $\Delta t_{obs}$ is
\begin{equation}
N^{atm,\nu_\mu}_{ev}[E^{\rm dep}_{\mu}\geq 10^4 \; {\rm GeV})]\simeq
\, 2 \times 10^{-8}\;\; \frac{\Delta t_{obs}}{min}
\end{equation}
So as long as the observed time window is at most of the order of
months the atmospheric background can be safely neglected even if one
of the events has deposited energy as low as $10^4$ GeV.

In summary revealing Planck-scale physics by comparing the arrival
times of muon track events from a single GRB at 
IceCube may be possible, but will be challenging. 

Observing the less frequent $\nu_e$ and $\nu_\tau$ events could
present us with a solution to the energy measurement problem. Because
of oscillations over cosmic distances the GRB flux will be equally
distributed over the three neutrino flavors. Neutrino telescopes
detect the Cherenkov light radiated by secondary particle showers
produced by neutrinos of all flavors. These include the
electromagnetic and hadronic showers initiated by $\nu_e$ and
$\nu_\tau$ as well as by neutral current interactions of neutrinos of
all flavors. Because the size of these showers, of order 10\,m in ice,
is small compared to the spacing of the PMTs, they represent, to a
good approximation, a point source of Cherenkov photons radiated by
the shower particles. These trigger the PMTs at the single
photoelectron level over essentially spherical volume, slightly
elongated in the direction of the initial neutrino, whose radius
scales linearly with the shower energy\cite{Halzen}.

Whereas the smaller first-generation telescopes mostly exploit the
large range of the muon to increase their effective area for
$\nu_{\mu}$, kilometer-scale detectors can fully exploit the
advantages associated with the detection of showers initiated by
$\nu_e$ and $\nu_{\tau}$:
\begin{enumerate}
\item They are detected over both Northern and Southern hemispheres.
\item The background of atmospheric neutrinos is significantly
reduced. At higher energies the muons from $\pi$ decay, the source of
atmospheric $\nu_{e}$, no longer decay and relatively rare K-decays
become the dominant source of background electron neutrinos.
\item $\nu_{\tau}$ are not absorbed, but degraded by energy in the
earth.
\end{enumerate}

Critical in this context is that, depending on energy, electron and
tau neutrinos deposit 0.5-0.8\% of their energy into an
electromagnetic shower initiated by the leading final state
lepton. The rest of the energy goes into the fragments of the target
that produce a second subdominant shower. For ice, the Cherenkov light
generated by shower particles spreads over a volume of radius 130\,m
at 10\,TeV and 460\,m at 10\,EeV, i.e. the shower radius grows by just
over 50\,m per decade in energy. The measurement of the radius of the
lightpool mapped by the lattice of PMTs determines the energy and
turns neutrino telescopes into total absorption calorimeters. Note
that even a contained ``direct hit" by a 10 EeV neutrino will not
saturate a km$^3$ detector volume. So, even for EeV neutrinos, IceCube
will not saturate and their energy spectrum can be measured.

The drawback of the use of shower events are the lower statistics and
the worse angular resolution.  Although, in principle, GRBs represent a
background-free signal in neutrino telescopes because of the
clustering in time and in direction, the event rates are still
uncertain. They will exceed 10 events per kilometer square year, in
the case where the baryon loading in GRB fireballs accommodates the
observed cosmic rays. With the low event rates anticipated doing the
Planck science with showers may not be a practical solution.
Furthermore their worse pointing will also difficulty background
rejection.

In the end we will have to await the actual observations to draw any
conclusions regarding the reality of observing Planck physics with
neutrino telescopes. We have argued that the existing
back-of-the-envelope estimates are unlikely to be realized and have
described the realistic background on which the possibility of
observing Planck physics will have to be pursued. The problem is
undoubtedly important given the absence of clear alternatives.

\acknowledgments

The authors thank the CERN theory group for its hospitality
during the early stages of this work. We thank Luis Anchordoqui for a careful reading of the manuscript. MCG-G is supported by MEC (Spain) Grant FPA-2004-00996 and 
by  National Science Foundation grant PHY-0354776.  
F.H. is supported in part by the National Science Foundation
grant PHY0098527, in part by the U.S.~Department of Energy under Grant
No.~DE-FG02-95ER40896, and in part by the University of Wisconsin
Research Committee with funds granted by the Wisconsin Alumni Research
Foundation.

\end{document}